\documentclass[twocolumn,aps,prl,superscriptaddress,nofootinbib,amsmath,amssymb,floatfix]{revtex4-2}

\usepackage[T1]{fontenc}
\usepackage{mathptmx}
\usepackage{mathrsfs}
\usepackage{bm}
\usepackage{graphicx}
\usepackage[dvipsnames]{xcolor}
\usepackage{tikz}
\usetikzlibrary{arrows.meta,positioning,calc}
\usepackage[colorlinks,citecolor=blue,linkcolor=blue,urlcolor=blue]{hyperref}
\usepackage{appendix}

\begin{document}

\newcommand{\be}{\begin{equation}}
\newcommand{\ee}{\end{equation}}
\newcommand{\bs}{\begin{split}}
\newcommand{\es}{\end{split}}
\newcommand{\hc}{\mathrm{h.c.}}
\newcommand{\Tr}{\mathrm{Tr}}
\newcommand{\Simp}{S_{xx}^{\rm imp}}
\newcommand{\ain}{\hat a_{\rm in}}
\newcommand{\bin}{\hat b_{\rm in}}
\newcommand{\cin}{\hat c_{\rm in}}
\newcommand{\R}[1]{\textcolor{red}{#1}}
\newcommand{\B}[1]{\textcolor{blue}{#1}}

\title{Exceptional-point Sensing with Quantum Advantage}

\author{Chuming Wang}
\affiliation{National Gravitational Laboratory, Hubei Key Laboratory of Gravitation and Quantum Physics, School of Physics, Huazhong University of Science and Technology, Wuhan, 430074, China}
\author{Xinyao Guo}
\affiliation{State Key Laboratory of Low Dimensional Quantum Physics,
Department of Physics, Tsinghua University, Beijing, 100084, China}
\author{Dezhi Zhang}
\affiliation{State Key Laboratory of Low Dimensional Quantum Physics,
Department of Physics, Tsinghua University, Beijing, 100084, China}
\author{Mingchen Liu}
\affiliation{State Key Laboratory of Low Dimensional Quantum Physics,
Department of Physics, Tsinghua University, Beijing, 100084, China}
\author{Yifei Li}
\affiliation{State Key Laboratory of Low Dimensional Quantum Physics,
Department of Physics, Tsinghua University, Beijing, 100084, China}
\author{Haixing Miao}
\email{haixing@tsinghua.edu.cn}
\affiliation{State Key Laboratory of Low Dimensional Quantum Physics,
Department of Physics, Tsinghua University, Beijing, 100084, China}
\author{Yiqiu Ma}
\email{myqphy@hust.edu.cn}
\affiliation{National Gravitational Laboratory, Hubei Key Laboratory of Gravitation and Quantum Physics, School of Physics, Huazhong University of Science and Technology, Wuhan, 430074, China}

\begin{abstract}
Exceptional-point sensors exploit singular non-Hermitian dynamics to enhance their response to weak perturbations. Recently, however, a no-go result has been established for a gain-loss-balanced dimer coupled to independent, phase-insensitive reservoirs: the exceptional-point susceptibility amplifies the signal and reservoir noise equally, precluding any quantum advantage. Here, we circumvent this limitation by engineering non-Hermitian dynamics in particle-hole Nambu space. A single common reservoir acts on modes forming a quantum-mechanics-free subsystem and generates an enhanced response protected from additional noise. We demonstrate quantum-metrological advantage for weak-signal sensing and static-parameter estimation. A continuous interpolation of the operator basis at fixed exceptional-point drift, with a singular midpoint, reveals how reservoir noise vanishes toward the protected limit.
Our results provide a route to exceptional-point sensing with genuine quantum-metrological advantage through reservoir engineering.
\end{abstract}

\maketitle

\emph{Introduction---}
Exceptional points (EPs) are non-Hermitian degeneracies at which eigenvalues and eigenvectors coalesce \cite{MiriAlu2019}. Near a second-order EP, a generic perturbation produces a square-root resonance splitting, motivating sensing proposals and experiments on different platforms \cite{Wiersig2014, Chen2017, Luo2022PseudoAntiPT, Wiersig2023Petermann,Sun2024Broadside, Bai2024NonlinearEP, Hu2025AntiPT,Ruan2025LossEnhanced,  Chen2026PhaseSensing, Almanakly2026QuantumEP}. Enhanced spectral response, however, does not by itself imply improved measurement precision \cite{Langbein2018,LauClerk2018,ChenJinLiu2019}: an effective non-Hermitian equation of motion is only the reduced description of a larger open quantum system. The physical implementation of gain, loss, drive, and readout also determines the quantum fluctuations against which the signal must be resolved.

Several studies have shown that the enhanced response near an EP can be accompanied by a corresponding enhancement of quantum noise \cite{LauClerk2018,ZhangSweeneyHsuYangStoneJiang2019,ChenJinLiu2019,
WangPetermann2020, Wang2025QuadraturePT, Wang2026SqueezingEP, WiersigRotter2026, WangEtAl2026HiddenSymmetry}. Loughlin and Sudhir made this limitation explicit for a gain-loss-balanced dimer \cite{LoughlinSudhir2024}: noise from its independent gain and loss reservoirs is amplified by the same EP susceptibility as the signal, eliminating the apparent quantum-limited advantage. 

Here, we show that this no-go result can be circumvented using a single engineered reservoir coupled to particle-hole Nambu modes that form a quantum-mechanics-free subsystem (QMFS) \cite{TsangCaves2012,WoolleyClerk2013,OckeloenKorppi2016,deLepinay2021}. The engineered reservoir generates the EP-enhanced signal response without concomitantly amplifying the reservoir noise: the correlated Langevin forces cancel exactly in the measured collective observable while adding in its conjugate partner. We demonstrate the resulting quantum advantage for both weak-signal sensing and static-parameter estimation, establish its equivalence to the auxiliary-cavity realizations of Refs.~\cite{Li2020,Li2021,Wang2022, Chen_Shu2022, Chen2025Cavities}, and identify the general Hamiltonian-level principle underlying this construction. This provides a genuinely dissipative route to EP-enhanced quantum sensing. It differs from coherent QMFS emulation without engineered dissipation \cite{WangClerk2019}; a recent four-mode superconducting implementation demonstrated enhanced sensitivity in a non-quantum-limited regime \cite{Assouly2026}.

\emph{Gain-loss-balanced dimer---} We first give a brief review of Loughlin and Sudhir's no go theorem for a gain-loss-balanced dimer with independent, phase-insensitive reservoir and focus on the weak-signal measurement. The non-Hermitian dynamics for the dimer $(a, b)$ is
\begin{align}\label{eq:a}
\dot a &= -\gamma a + g b\,,\\
\dot b &= \gamma b - g a\,. \label{eq:b}
\end{align}
The normal mode frequencies 
are $\pm\sqrt{g^2-\gamma^2}$ and coalesce at the EP with $g=\gamma$. The corresponding  collective EP eigenmode $d_{-}$ is the difference of $a$ and $b$, namley, 
\be
d_{-}\equiv (a-b)/\sqrt{2}\,.
\ee
A minimal Hamiltonian embedding for such non-Hermitian dynamics requires a coherent coupling between $a$ and $b$, promoting the amplitudes to the quantum operators, 
\be
    \hat H_{c}
    = i\hbar g(\hat a^\dagger\hat b-\hat a\hat b^\dagger)\,,
    \label{eq:H_LS}
\ee
and also couplings to two independent reservoirs with input noise operators $\hat a_{\rm in}$ and $\hat b_{\rm in}$, respectively. The resulting  quantum Langevin 
equations of motion are given by
\begin{align}\label{eq:hata}
    \dot{\hat a}
    &=-\gamma\hat a+g\hat b+\sqrt{2\gamma}\,\ain
    +i\sqrt{2}\alpha x,\\
    \dot{\hat b}
    &=\gamma\hat b-g\hat a+\sqrt{2\gamma}\,\bin^\dagger\,,
    \label{eq:hatb}
\end{align}
where we have included a weak signal $x$ coupled to the mode $a$ via a coupling strength of $\alpha$: $\hat H_x=-\sqrt2\hbar\alpha x(t)(\hat a+\hat a^\dagger)$. 

At the EP with $g=\gamma$, the collective EP mode $\hat d_{-}$ obeys $
\dot {\hat d}_-=i\alpha x+\sqrt{\gamma}(\ain-\bin^\dagger)$, which, in the frequency domain, reads
\be
\label{eq:uLS}
{\hat d}_-(\Omega) = -\frac{1}{i\Omega}\left\{ i\alpha x(\Omega) + \sqrt{\gamma}[\hat a_{\rm in}(\Omega)-\hat b^{\dagger}_{\rm in}(-\Omega)]\right\}
\ee
The first term in the bracket is the signal term while the second one is the noise term. Both terms are multiplied by $\Omega^{-1}$, which is  divergent at the EP eigen-frequency $\Omega=0$. This shared amplification factor for both terms is the no-go mechanism: increasing the EP susceptibility amplifies both the signal and the quantum noise, preventing any enhancement of the signal-to-noise ratio in the quantum-limited regime.

\begin{figure}[!t]
\centering
\begin{tikzpicture}[
    scale=1.2,                         
    every node/.style={transform shape}, 
    font=\scriptsize,
    title/.style={draw, rounded corners=2pt, thick, fill=gray!8, align=center, 
                  inner xsep=10pt, inner ysep=4pt},
    panel/.style={draw, rounded corners=2pt, thick},
    mode/.style={circle, draw, thick, minimum size=6.0mm, inner sep=0pt},
    bath/.style={draw, rounded corners=1.2pt, fill=gray!10, align=center, inner xsep=2pt, inner ysep=1.4pt},
    outcome/.style={draw, rounded corners=1.5pt, thick, align=center, inner xsep=4pt, inner ysep=2pt},
    arr/.style={-{Latex[length=1.5mm]}, thick},
    link/.style={-{Latex[length=1.5mm]}, thick},
    noi/.style={-{Latex[length=1.6mm]}, thick, dashed, red!70!black}
]
    \node[title, font=\small] (top) at (3.05,3.55) {EP sensor};

    \draw[panel] (-0.15,0.85) rectangle (2.60,2.75);
    \draw[panel] (3.50,0.85) rectangle (6.25,2.75);
    \node[font=\footnotesize, align=center] (lslabel) at (1.23,2.95) {gain-loss-balanced dimer};
    \node[font=\footnotesize, align=center] (cblabel) at (4.88,2.95) {noise-protected mode};
    \draw[link] (top.south west) -- (lslabel.north);
    \draw[link] (top.south east) -- (cblabel.north);

    \node[mode, fill=Blue!7] (la) at (0.65,1.82) {$\hat a$};
    \node[mode, fill=Orange!12] (lb) at (1.85,1.82) {$\hat b$};
    \draw[{Latex[length=1.25mm]}-{Latex[length=1.25mm]}, thick] (la) -- node[above] {$g$} (lb);
    \node[bath] (lain) at (0.23,2.50) {$a_{\rm in}$};
    \node[bath] (laout) at (0.23,1.10) {$a_{\rm out}$};
    \node[bath] (lbin) at (2.27,2.50) {$b_{\rm in}^\dagger$};
    \node[bath] (lbout) at (2.27,1.10) {$b_{\rm out}$};
    \draw[noi] (lain) -- (la);
    \draw[noi] (lbin) -- (lb);
    \draw[arr] (la) -- (laout);
    \draw[arr] (lb) -- (lbout);

    \node[mode, fill=Blue!7] (ra) at (4.275,2.20) {$\hat a$};
    \node[mode, fill=Orange!12] (rb) at (5.475,2.20) {$\hat b^{\dag}$};
    \node[circle, fill=black, minimum size=1.8mm, inner sep=0pt] (bullet) at (4.875,1.60) {};
    \draw[{Latex[length=1.5mm]}-{Latex[length=1.5mm]}, thick] (bullet) -- node[left] {$g$} (ra);
    \draw[{Latex[length=1.5mm]}-{Latex[length=1.5mm]}, thick] (bullet) -- node[right] {$g$} (rb);
    \node[bath, fill=white] (c_in) at (4.275,1.05) {$c_{\rm in}$};
    \node[bath, fill=white] (c_out) at (5.475,1.05) {$c_{\rm out}$};
    \draw[noi] (c_in) -- (bullet);
    \draw[noi] (bullet) -- (c_out);

    \node[outcome, fill=red!5, text=red!70!black] (noadv) at (1.30,0.30)
        {no quantum advantage};
    \node[outcome, fill=Green!6, text=Green!45!black] (adv) at (4.875,0.30)
        {quantum advantage};
    \draw[link] (1.30,0.85) -- (noadv.north);
    \draw[link] (4.875,0.85) -- (adv.north);
\end{tikzpicture}
\caption{Hamiltonian embeddings for the gain-loss-balanced dimer (left) and the noise-protected EP mode (right). For the latter, one common reservoir with input/output fields is coupled to $\hat a$ and $\hat b$; the common input noise cancels from the EP mode $\hat u=(\hat a-\hat b^\dagger)/\sqrt2$, allowing the EP response to yield a quantum advantage.}
\label{fig:implementation_compare}
\end{figure}
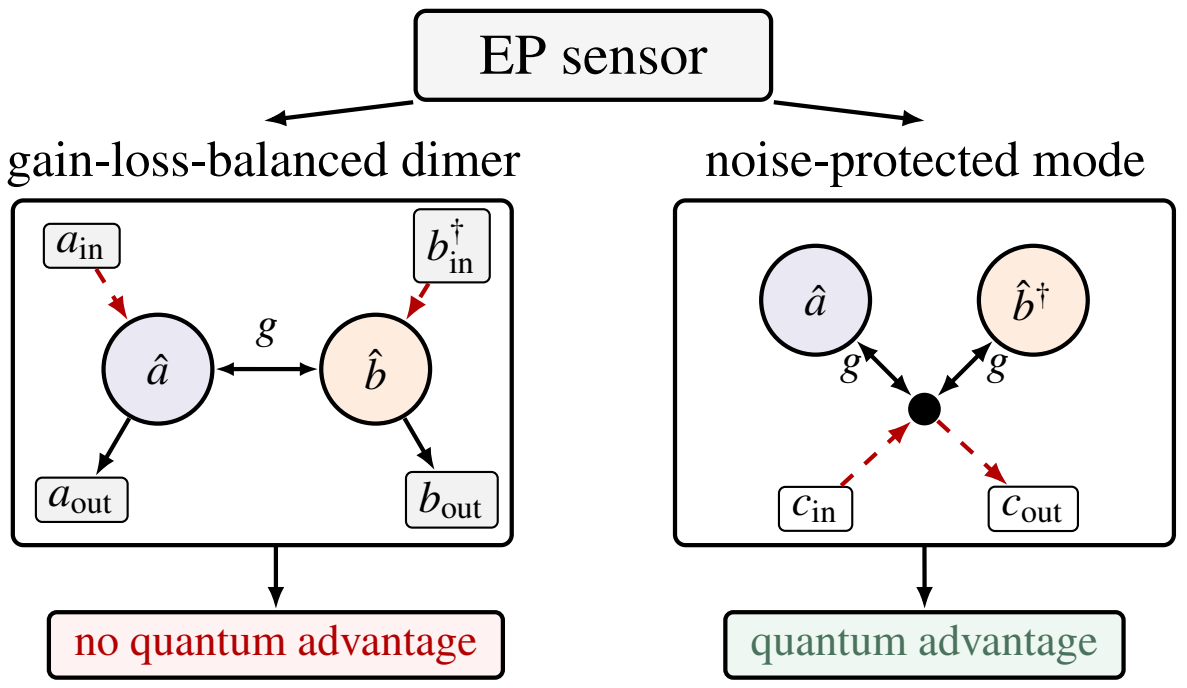

\emph{Non-Hermitian dynamics in Nambu space---} The Hamiltonian embedding of the non-Hermitian dynamics is drastically different if we consider the particle-hole Nambu space $(\hat a, \hat b^{\dag})$ instead of the space spanned by $(\hat a, \hat b)$. As illustrated in Fig.\,\ref{fig:implementation_compare}, the minimal Hamiltonian embedding for $(\hat a, \hat b^{\dag})$ only requires a common reservoir coupled to the EP mode $\hat u=(\hat a-\hat b^{\dag})/\sqrt{2}$, without the need of additional coherent coupling between $\hat a$ and $\hat b$ (the derivation is given in Appendix A). The corresponding Langevin equations are
\begin{align}
\label{eq:eom_minimal_cb1}
    \dot{\hat a}
    &=-\gamma\hat a+\gamma\hat b^\dagger
+\sqrt{2\gamma}\,\cin+i\sqrt2\alpha x(t),\\
    \dot{\hat b}^\dagger
    &=\gamma\hat b^\dagger
    -\gamma\hat a+\sqrt{2\gamma}\,\cin.\label{eq:eom_minimal_cb2}
\end{align}
The homogeneous part of the above equation again has a second-order EP: the two eigenfrequencies coalesce at $0$, and only one normal mode remains independent. 

In contrast to Eq.\,\eqref{eq:hata} and Eq.\,\eqref{eq:hatb}, both $\hat a$ and $\hat b^{\dag}$ are driven by the same noise operator $\hat c_{\rm in}$ from the common reservoir. Therefore, the new EP mode in the Nambu space, defined as 
\be
\label{eq:u}
\hat u\equiv (\hat a-\hat b^\dagger)/\sqrt{2}\,,
\ee 
satisfies $\dot{\hat u}=i\alpha x$, which in the frequency domain is
\be
\label{eq:unew}
\hat u(\Omega)= -\frac{\alpha}{\Omega}x(\Omega)\,.
\ee
Therefore, such a quantum-noise-protected EP mode has a divergent signal response at the EP pole without additional noise.  In this example, this ideal divergence at $\Omega=\Omega_{EP}=0$ is regularized by finite time, residual loss, and  nonlinearities, which are practical imperfections rather than the fundamental limitation that amplifies both signal and noise in the gain-loss-balanced dimer discussed earlier and also in Ref.~\cite{LoughlinSudhir2024}


\emph{Quantum advantage for weak-signal detection---}
\label{sec:snr} Eq.\,\eqref{eq:uLS} and Eq.\,\eqref{eq:unew} already show the EP mode defined in the Nambu space demonstrates quantum-noise protection compared to the one for the gain-loss-balanced dimmer. We now show explicitly that such quantum advantage persists when we finally read out the outgoing field of the reservoir. 
According to the standard input-output relation\,\cite{GardinerCollett1985}, we have
\be
\hat c_{\rm out}=\hat c_{\rm in}- \sqrt{2\gamma}\,(\hat a-\hat b^{\dag}) = \hat c_{\rm in} - 2\sqrt{\gamma}\, \hat u\,.
\label{eq:io}
\ee
Using the homodyne detection to measure the phase quadrature of outgoing field, defined as
$\hat Y_{c{\rm out}}(\Omega) \equiv [\hat c_{\rm out}(\Omega)-\hat c_{\rm out}(-\Omega)^\dagger]/(\sqrt{2}i)$ in the frequency domain, we obtain
\be
    \hat Y_{c{\rm out}}(\Omega)
    =\hat Y_{c{\rm in}}(\Omega)-\frac{2\sqrt{2\gamma}\alpha}{i\Omega}\,x(\Omega).
    \label{eq:Yout}
\ee
For vacuum input noise, $S^{\rm in}_{YY}(\Omega)=1$ in the single-sided convention\,\cite{Braginsky1992QuantumMeasurement, ClerkEtAl2010}, and therefore the weak-signal measurement imprecision, normalized by the signal response, is
\be
    \Simp(\Omega)
    =\frac{\Omega^2}{8\gamma \alpha^2}\,.
    \label{eq:Simp}
\ee
The imprecision vanishes as $\Omega$ approaches $\Omega_{\rm EP}=0$ instead of being flatten out for the gain-loss-balanced dimer. 

\emph{Quantum advantage for parameter estimation---}We now consider the quantum advantage for parameter estimation in the Nambu space. Let a small  parameter $\epsilon>0$ to be estimated perturbs the off-diagonal coupling:
\begin{align}
    \dot{\hat a}
    &\approx -\gamma\hat a+\gamma(1+\epsilon)\hat b^\dagger
+\sqrt{2\gamma}\,\cin,\label{eq:eom_minimal_cb1}\\
    \dot{\hat b}^\dagger
    &\approx \gamma\hat b^\dagger
    -\gamma(1+\epsilon)\hat a+\sqrt{2\gamma}\,\cin.\label{eq:eom_minimal_cb2}
\end{align}
Here we have neglected additional noise terms of the order $\sqrt{\epsilon}$, which does not change the conclusion on the quantum advantage (details are provided in Appendix B). 
Introducing 
\be
\label{eq:v}
\hat v\equiv (\hat a+\hat b^{\dag})/\sqrt{2}\,
\ee
which is conjugate to $\hat u$, the above equations can be rewritten as 
\begin{align}
    \dot{\hat u}&=\gamma\epsilon\,\hat v,\\
    \dot{\hat v}&=-\gamma(2+\epsilon)\hat u
    +2\sqrt{\gamma}\hat c_{\rm in}.\label{eq:static_apm}
\end{align}
Eliminating $\hat v$ leads to 
\be
    \ddot {\hat u}+
    \Omega_\epsilon^2{\hat u}=2\gamma^{3/2}\epsilon\,\hat c_{\rm in}\,,\label{eq:static_second_order}
\ee
where we have defined $\Omega_\epsilon\equiv \gamma\sqrt{\epsilon(2+\epsilon)}$.  The resonant frequencies 
are $\Omega_\pm=\pm\Omega_\epsilon$ which shows the standard square-root splitting common for EP sensors. Togehter with Eq.\,\eqref{eq:io}, the resulting noise spectrum for the output phase quadrature of $\hat c_{\rm out}$ around $\Omega=\Omega_\pm+\Delta\omega$ is
\be
   S_{YY}^{\rm out}(\Omega_\pm+\Delta\omega)
    \approx \frac{2\gamma^2\epsilon}{\Delta\omega^2}\,,
\label{eq:static_output_noise}
\ee
where we have assumed that the frequency of interest is sufficiently close to $\Omega_{\pm}$ and $\epsilon\gamma^2/\Delta\omega^2\gg 1$ in the approximation. It can be converted into the following frequency noise spectrum\,\cite{LoughlinSudhir2024}, given a coherent output amplitude $\alpha_\pm$, 
\be
S_{\dot\varphi\dot\varphi}
(\Omega_\pm+\Delta\omega)
    =
    \frac{\Delta\omega_\pm^2}{4\gamma|\alpha_\pm|^2}
    S_{YY}^{\rm out}
    \simeq
    \frac{\gamma\,\epsilon} {2|\alpha_\pm|^2}.
    \label{eq:static_freq_noise}
\ee
If we measure the frequency splitting, the sensitivity is given by 
\be
    \mathcal S
    =
 \left|\frac{\partial(\Omega_+-\Omega_-)}
 {\partial\epsilon}\right|
    =
\frac{2\gamma(1+\epsilon)}
{\sqrt{\epsilon(2+\epsilon)}}
    \propto \epsilon^{-1/2}\,. 
\label{eq:static_sensitivity}
\ee
The resulting parameter-estimation imprecision  $\epsilon_{\rm imp}\propto\sqrt{S_{\dot\varphi\dot\varphi}}/\mathcal S$, scales as
\be
    \epsilon_{\rm imp}
    \propto \epsilon.
\label{eq:static_imprecision_scaling}
\ee
This shows the quantum advantage for the EP sensing, which is in contrast to the gain-loss-balanced dimer as shown in Ref.\,\cite{LoughlinSudhir2024}, the imprecision is independent of $\epsilon$, namely,  
\be
\epsilon^{\rm EP}_{\rm imp}
\propto \epsilon^0\,, 
\ee  
which is because the near-resonant frequency noise spectrum amplitude is proportional to $\epsilon^{-1/2}$ and exactly cancels the EP sensitivity scaling. 

\begin{figure}[!t]
\centering
\begin{tikzpicture}[
    scale=1.4,                         
    every node/.style={transform shape}, 
    font=\scriptsize,
    box/.style={draw, rounded corners=2pt, thick, align=center, inner sep=3pt, minimum width=21mm, minimum height=8mm},
    bath/.style={draw, thick, align=center, inner sep=2pt, minimum width=16mm, minimum height=7mm},
    arr/.style={-{Latex[length=1.8mm]}, thick},
    darr/.style={-{Latex[length=1.8mm]}, thick, dashed},
    noarr/.style={-{Latex[length=1.8mm]}, thick, red!70!black}
]
    \node[box, fill=Blue!7] (u) {noise-protected \\EP mode $\hat{\bm u}$};
    \node[box, fill=Orange!12, right=17mm of u] (v) {conjugate\\mode $\hat{\bm v}$};
    \node[bath, fill=gray!12, below=8mm of $(u)!0.5!(v)$] (bath) {common\\reservoir};
    \node[bath, fill=Green!10, above=8mm of u] (sig) {signal};
    \node[bath, fill=Purple!10, above=8mm of v] (out) {output};
    \draw[arr] (sig) -- node[left] {$x$} (u);
    \draw[arr] (u) -- node[above] {one-way} (v);
    \draw[arr] (v) to[bend left=18] node[above] {$\epsilon$} (u);
    \draw[arr] (bath) -- node[right=3pt,yshift=-3pt] {input noise} (v);
    \draw[arr] (u) -- node[above, sloped] {readout} (out);
    \draw[darr] (bath) -- ($(u.south)+(0,-1mm)$);
    \node[red!70!black, left=0mm of bath.west, align=right] {no direct\\noise drive};
\end{tikzpicture}
\caption{Illustration of the working principle for a quantum-noise protected EP sensor.  The measured EP variable $\hat{\bm u}$ carries the signal, while common-reservoir input noise is routed to the conjugate sector $\hat{\bm v}$.  The one-way coupling $\hat{\bm u}\to\hat{\bm v}$ protects $\hat u$ from the input noise; a perturbation $\epsilon$ closes the reverse path and produces the square-root mode frequency splitting.}
\label{fig:principle}
\end{figure}
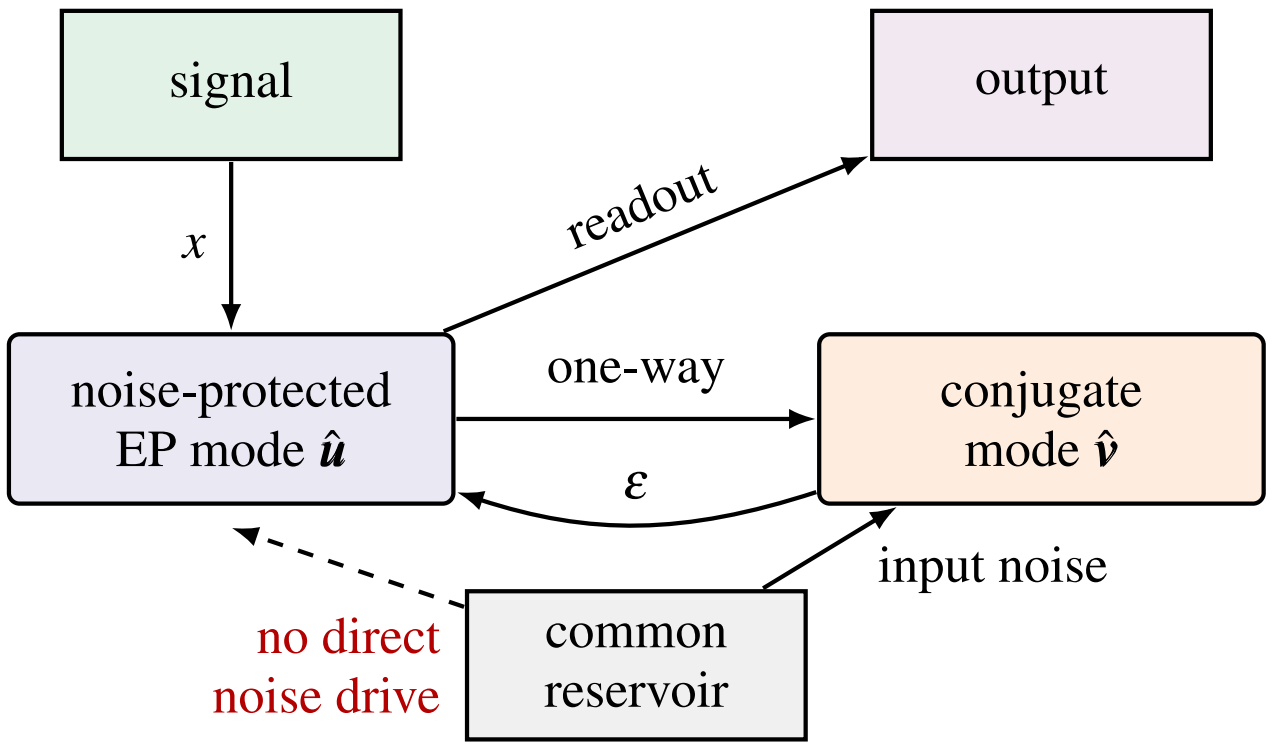

\emph{Working principle---}
\label{sec:principle}
The quantum advantage discussed above is intimately related to QND measurements with QMFS, which underlies its working principle. This can be better understood by considering the EP mode $\hat u$ and its conjugate pair $\hat v$, which we defined in Eq.\,\eqref{eq:u} and Eq.\,\eqref{eq:v}, respectively. They satisfy
  $  [\hat u,\,\hat u^\dagger]=0,
    [\hat u,\,\hat v^\dagger]=1$.
The continuum field of the common reservoir for $\hat a$ and $\hat b^{\dag}$ is only coupled to $\hat u$ via $\hat H_{\rm ext}$, so that we have $[\hat u,\, \hat H_{\rm ext}]=0$. Since there is no coherent coupling between $\hat u$ and $\hat v$ and no free Hamiltonian for $\hat v$, $\hat u$ is commuting with itself at different times, namely, 
\be
[\hat u(t),\,\hat u^{\dag}(t')]=0\,.
\ee
Therefore, the Hermitian quadratures of $\hat u$: $\hat X_u\equiv (\hat u+\hat u^{\dag})/\sqrt{2}$ and $\hat Y_u\equiv (\hat u-\hat u^{\dag})/(\sqrt{2}i)$  are indeed (QND) observables which form the QMFS\,\cite{TsangCaves2012}. 
As illustrated in Fig.\,\ref{fig:principle}, such a Hamiltonian structure also means that the coupling of $\hat u$ to the common reservoir not only induces a one-way coupling from $\hat u$ to $\hat v$ since $\hat U(t)\hat u\hat U^\dag(t)$ has no dependence on $\hat v$. While $\hat U(t)\hat u\hat U^\dag(t)$ will contain $\hat v$ since $[\hat u,\hat v^\dagger]=1$, which means the input noise from the reservoir is only driving $\hat v$ without directly driving $\hat u$. This makes the EP mode $\hat u$ protected from quantum noise of the reservoir, which is the essence of noise-protected EP sensing.

\emph{Minimal realization of a common reservoir---} The Hamiltonian embedding for $(\hat a,\,\hat b^{\dag})$ can be realized by coherently coupling the EP mode $\hat u$ to an auxiliary mode $\hat c$ with a fast dynamics: 
\be
    \hat H_{\rm aux}
    =i\sqrt{2}\hbar 
    g_c (\hat u\hat c^\dagger-\hat u^{\dag}\hat c),
    \label{eq:H_cb}
\ee
and $\hat c$ is in turn coupled to the external continuum field which defines $\hat c_{\rm in}$ and $\hat c_{\rm out}$. 
The corresponding equations of motion for $\hat a$ and $\hat b^{\dag}$ are
\begin{align}
\dot {\hat a}& =-g_c \hat c\,,\quad \dot {\hat b}^{\dag} =-g_c \hat c\,,
\\
\dot {\hat c}& =-\kappa \hat c+ g_c(\hat a-\hat b^{\dag})-\sqrt{2\kappa}\,\hat c_{\rm in}\,.
\label{eq:c}
\end{align}
Adiabatically eliminating the fast auxiliary mode $\hat c$ by ignoring $\dot{\hat c}$ in Eq.\,\eqref{eq:c}, we have $\hat c=g_c(\hat a-\hat b^{\dag})/\kappa+\sqrt{2/\kappa} \hat c_{\rm in}$, leading to
\begin{align}
\dot {\hat a} & = -\gamma_{\rm eff} \hat a + \gamma_{\rm eff}\hat b^{\dag} + \sqrt{2\gamma_{\rm eff}}\,\hat c_{\rm in}\,,\\
\dot {\hat b}^{\dag} & = \gamma_{\rm eff} \hat b^{\dag} - \gamma_{\rm eff} \hat a +\sqrt{2\gamma_{\rm eff}}\,\hat c_{\rm in}\,,
\end{align}
where $\gamma_{\rm eff} \equiv {g_c^2}/{\kappa}$. Identifying $\gamma_{\rm eff}$ with $\gamma$, they recover the equations of motion shown in Eq.\,\eqref{eq:eom_minimal_cb1} and Eq.\,\eqref{eq:eom_minimal_cb2}.Such a PT-symmetric EP sensor model was first proposed in Refs.~\cite{Li2020,Li2021,Wang2022, Chen_Shu2022,Chen2025Cavities}, which was motivated by reducing the quantum noise of gravitational-wave detectors and dark matter search. This has also been the main mechanism for realising dissipative pairing interactions
in bosonic lattice\,\cite{Pocklington2023}. 

\emph{Interpolating between Hamiltonian embeddings---}
Interestingly, there is a continuous interpolation between the gain-loss-balanced dimer and the Nambu-space model via a fixed EP drift by continuously varying the operator basis. For $0\leq\theta\leq\pi/2$, define
\be
  \hat u_\theta=(\hat a-\hat b_\theta)/\sqrt2,\quad \hat b_\theta\equiv \hat b\cos\theta+\hat b^\dagger\sin\theta.
\label{eq:embedding_basis}
\ee
Thus $\hat u_0=\hat d_-$ in the gain-loss-balanced dimer and $\hat u_{\pi/2}=\hat u$ in the Nambu-space model. With the signal temporarily omitted, impose the equations of motion:
\be
    \dot{\hat{\bm z}}_\theta=\bm{M}\hat{\bm z}_\theta
    +\bm{B}_\theta\hat{\bm\xi}_{\rm in},\qquad
    M=\gamma\begin{pmatrix}-1&1\\-1&1\end{pmatrix},
\label{eq:embedding_drift}
\ee
where $\hat{\bm z}_\theta=(\hat a,\hat b_\theta)^T$ and $\hat{\bm\xi}_{\rm in}=(\hat\xi_+,\hat\xi_-^\dagger)^T$ are uncorrelated vacuum inputs, each with single-sided quadrature spectrum one. Since $M^2=0$ and $M\neq0$, the system dynamics retains the same second-order EP throughout. The interpolation is not canonical: $[\hat b_\theta,\hat b_\theta^\dagger]=k\equiv\cos2\theta$, so the commutator matrix $(\bm{G}_\theta)_{ij}\equiv[\hat{\bm z}_{\theta i},(\hat{\bm z}_{\theta j})^\dag]$ is $\bm{G}_\theta=\operatorname{diag}(1,k)$ and $[\hat u_\theta,\hat u_\theta^\dagger]=(1+k)/2=\cos^2\theta$. The EP variable therefore becomes a QMFS variable only at the particle-hole endpoint with $\theta=\pi/2$. 

For canonical input operators $\hat{\bm\xi}_{\rm in}$ with commutator matrix $\bm{J}=\operatorname{diag}(1,-1)$, preservation of $\bm{G}_\theta$ requires
\be
\begin{aligned}
    \bm{B}_\theta \bm{J} \bm{B}_\theta^\dagger
    &=-\bm{M}\bm{G}_\theta-\bm{G}_\theta \bm{M}^\dagger=\gamma\begin{pmatrix}2&1-k\\1-k&-2k\end{pmatrix}\,,
\end{aligned}
\label{eq:embedding_realizability}
\ee
which fixes the noise commutator but does not completely fix $\bm{B}_\theta$.  One possible spectral factorization of the commutator matrix is
\be
    \bm{B}_\theta=\begin{pmatrix}
    \sqrt{r_+}\cos\phi&-\sqrt{r_-}\sin\phi\\
    \sqrt{r_+}\sin\phi&\sqrt{r_-}\cos\phi
    \end{pmatrix},
\label{eq:embedding_noise_matrix}
\ee
where $r_\pm=\gamma[\sqrt{2(1+k^2)}\pm(1-k)]$, and $\tan2\phi=(1-k)/(1+k)$ with $0\leq\phi\leq\pi/4$.

Restoring the weak signal sensing gives $\dot{\hat u}_\theta=i\alpha x+\hat F_\theta$, where $\hat F_\theta=(1,-1)B_\theta\hat{\bm\xi}_{\rm in}/\sqrt2$. In the single-sided vacuum convention used above, the phase-quadrature force $\hat Y_F=(\hat F_\theta-\hat F_\theta^\dagger)/(\sqrt2 i)$ has spectrum
\be
    S_{Y_FY_F}^{(\theta)}
    =\frac{\gamma(1+k)^2}{d}
    =\frac{4\gamma\cos^4\theta}
    {\sqrt{2(1+\cos^2 2\theta)}}.
\label{eq:embedding_noise_spectrum}
\ee
It decreases monotonically from $2\gamma$ to zero, vanishing as $2\gamma(\pi/2-\theta)^4$ near the protected endpoint. Thus the reservoir force is suppressed while the EP susceptibility and signal coupling remain fixed. This spectrum characterizes the chosen reservoir realization, rather than a universal noise bound inferred from $M$ alone.

For $k\neq0$, an explicit input-output realization uses the coherent Hamiltonian
\be
    \hat H_\theta=i\hbar {\gamma(1+k)}/{(2k)}
    (\hat a^\dagger\hat b_\theta-\hat b_\theta^\dagger\hat a),
\label{eq:embedding_hamiltonian}
\ee
and two independent continuum fields of the reservior $\hat\xi_{\sigma,z}$, with $\sigma=\pm$ and $[\hat\xi_{\sigma,z},\hat\xi_{\sigma',z'}^\dagger]=\delta_{\sigma\sigma'}\delta(z-z')$. Each field has the free Hamiltonian and propagation convention of Appendix~\ref{app:embedding}. 
Define the system coupling combinations
\be
\begin{aligned}
    \hat s_{\theta,+}&=\sqrt{r_+}\left(
    \cos\phi\,\hat a+\frac{\sin\phi}{k}\hat b_\theta\right),\\
    \hat s_{\theta,-}&=\sqrt{r_-}\left(
    \sin\phi\,\hat a^\dagger
    -\frac{\cos\phi}{k}\hat b_\theta^\dagger\right), 
\end{aligned}
\label{eq:embedding_couplings}
\ee
and their corresponding interaction Hamiltonian with the continuum field is given by
\be
\hat H_{\theta,\rm ext}=i\hbar\sum_{\sigma=\pm}
\left(\hat\xi_{\sigma,0}\hat s_{\theta,\sigma}^\dagger
-\hat\xi_{\sigma,0}^\dagger\hat s_{\theta,\sigma}\right).
\label{eq:embedding_continuum}
\ee
 The Heisenberg equation for any system operator is
\begin{align}
\dot{\hat o}=\frac{i}{\hbar}[\hat H_\theta,\hat o]
+\sum_{\sigma=\pm}\Bigl(
[\hat o,\hat s_{\theta,\sigma}^\dagger]\hat\xi_{\sigma,0}
-\hat\xi_{\sigma,0}^\dagger[\hat o,\hat s_{\theta,\sigma}]\Bigr).
\label{eq:embedding_heisenberg}
\end{align}
With $\hat o=\hat a,\hat b_\theta$ and solution for the continuum fields, we arrive at Eq.~\eqref{eq:embedding_drift}, including exactly the noise matrix in Eq.~\eqref{eq:embedding_noise_matrix}. The sign of $\hat s_{\theta,-}$ fixes the reservoir phase to match that matrix.

The coherent term $\hat H_\theta$ combines beam-splitter and two-mode-squeezing interactions. At $\theta=0$, it reduces to Eq.~\eqref{eq:H_LS} with $g=\gamma$, while $\hat s_{\theta,+}=\sqrt{2\gamma}\hat a$ and $\hat s_{\theta,-}=-\sqrt{2\gamma}\hat b^\dagger$: the independent-reservoir dimer is recovered. At $\theta=\pi/2$, both $\hat H_\theta$ and $\hat s_{\theta,-}$ vanish, leaving $\hat s_{\theta,+}=\sqrt{2\gamma}(\hat a-\hat b^\dagger)=2\sqrt\gamma\hat u$. This recovers the common-reservior interaction; $[\hat u,\hat s_{\theta,+}]=[\hat u,\hat s_{\theta,+}^\dagger]=0$, so the reservoir force on $\hat u$ vanishes.

One subtlety arises when $\theta=\pi/4$. In this case, we have $\hat b_{\pi/4}=(\hat b+\hat b^\dagger)/\sqrt{2}$ is Hermitian, but the basis $(\hat a, \hat b_{\pi/4})$ discards its conjugate quadrature $(\hat b-\hat b^\dagger)/(i\sqrt{2})$. Then the commutation matrix $G_{\pi/4}=\mathrm{diag}(1,0)$ is singular, and imposing the same $M$ gives the non-Hermitian equation $\dot{\hat b}_{\pi/4}=-\gamma\hat a+\gamma\hat b_{\pi/4}$, which cannot be a physical quadrature equation. In addition, the divergences of $\hat H_\theta$ and $\hat s_{\theta,\pm}$ as $k\to0$ when $\theta=\pi/4$
are also the signatures for the failure of the reduced two-dimensional description $(\hat a, \hat b_{\pi/4})$---so a regular interpolation must retain missing conjugate quadrature and use an enlarged drift matrix. Away from this singularity, the explicit embeddings demonstrate how identical EP drift matrices can support a quantum-noise behavior with metrological advantages.

\emph{Discussion and conclusion---}
The Loughlin--Sudhir no-go theorem applies to a gain-loss-balanced EP dimer whose signal mode is directly driven by independent, phase-insensitive reservoir noise. The common-reservoir sensor shows that this is not a universal limitation of EP metrology: its EP variable forms a protected quantum-mechanics-free subspace, while reservoir noise drives only the conjugate partner. Earlier three-mode PT-symmetric proposals realize this model in the fast-auxiliary-mode limit. The fixed-drift comparison further identifies the Hamiltonian and reservoir embedding, together with the operator commutators, as essential to the quantum-noise behavior; the non-Hermitian drift alone does not determine the metrological performance. Practical imperfections such as finite observation time, internal loss, and nonlinear saturation will bound the achievable enhancement, but they are distinct from the noise projection responsible for the independent-reservoir no-go result.

\emph{Acknowledgement---}Y.\,M. and C.\,W is supported by the National Natural Science Foundation of China under Grant No.12474481 and No. 12441503, National Key R$\&$D Program of China (2023YFC2205801). X. G., D. Z., M. L., Y. L., and H. M. are supported by National Natural Science Foundation of China under Grant No. 12441503 and National Key R$\&$D Program of China (2023YFC2205800).

\setcounter{secnumdepth}{2}
\appendix
\section{Minimal continuum embedding}
\label{app:embedding}

Here we provide the details for the Langevin equations Eqs.~\eqref{eq:eom_minimal_cb1}--\eqref{eq:eom_minimal_cb2}. 
We use a continuum field with $[\hat c_z,\hat c_{z'}^\dagger]=\delta(z-z')$ and propagation speed set to unity. The local interaction and free-field Hamiltonians are \cite{GardinerCollett1985, Chen_2013}
\begin{align}
\hat H_{\rm ext}&=i\hbar g_B(\hat c_0\hat u^\dagger-\hat c_0^\dagger\hat u),\label{eq:continuum_H}\\
\hat H_{\rm con}&=\frac{i\hbar}{2}\int_{-\infty}^{\infty}dz
\left[(\partial_z\hat c_z)\hat c_z^\dagger-\mathrm{h.c.}\right].
\end{align}
Here $g_B$ is a continuum coupling, distinct from the coherent dimer coupling $g$. The field equation is
\be
(\partial_t-\partial_z)\hat c_z=-g_B\hat u\delta(z).
\ee
Integrating across the interaction point gives
\be
\hat c_{0-}=\hat c_{0+}-g_B\hat u.
\ee
With $\hat c_{0+}=\cin$, $\hat c_{0-}=\hat c_{\rm out}$, and the symmetric junction convention,
\be
\hat c_0=(\hat c_{0-}+\hat c_{0+})/2
=\cin-\tfrac12g_B\hat u.
\ee
The system equations consequently read
\begin{align}
\dot{\hat a}&=\frac{g_B}{\sqrt2}\cin-\frac{g_B^2}{2\sqrt2}\hat u,\\
\dot{\hat b}^\dagger&=\frac{g_B}{\sqrt2}\cin-\frac{g_B^2}{2\sqrt2}\hat u.
\end{align}
Setting $g_B=2\sqrt\gamma$ reproduces Eqs.~\eqref{eq:eom_minimal_cb1}--\eqref{eq:eom_minimal_cb2} without the signal, and $\hat c_{\rm out}=\cin-2\sqrt\gamma\hat u$. Adding $\hat H_x$ supplies the signal term. This normalization fixes both damping and readout consistently.

\section{Physical completion of the perturbed sensor and its noise scaling}
\label{app:noise}

In this appendix, we will show the exact Markovian input-output realization  of Eqs.\,\eqref{eq:eom_minimal_cb1} and \eqref{eq:eom_minimal_cb2} for $\epsilon>0$ without neglecting the additional noise term of the order of $\sqrt{\epsilon}$. Define
\begin{equation}
q=\gamma(2+\epsilon),\qquad r=\gamma\epsilon\,.
\end{equation}
Introduce independent continuum fields $\hat c_z$ and $\hat d_z$, each with the free Hamiltonian and propagation convention of Appendix~\ref{app:embedding}, and with vacuum inputs $\hat c_{\rm in}$ and $\hat d_{\rm in}$. Their equal-time commutators are $[\hat c_z,\hat c_{z'}^\dagger]=[\hat d_z,\hat d_{z'}^\dagger]=\delta(z-z')$, with all interchannel commutators zero. In terms of $\hat u=(\hat a-\hat b^\dagger)/\sqrt2$ and $\hat v=(\hat a+\hat b^\dagger)/\sqrt2$, choose the local interaction
\be
\hat H_{\rm ext}^{(\epsilon)}=i\hbar\sqrt{2q}
\left(\hat c_0\hat u^\dagger-\hat c_0^\dagger\hat u\right)
-i\hbar\sqrt{2r}
\left(\hat d_0\hat v-\hat d_0^\dagger\hat v^\dagger\right).
\label{eq:complete_continuum}
\ee
The relative sign fixes a reservoir phase convention. The equations of motion for the continuum fields are
\begin{align}
(\partial_t-\partial_z)\hat c_z&=-\sqrt{2q}\,\hat u\delta(z),\\
(\partial_t-\partial_z)\hat d_z&=+\sqrt{2r}\,\hat v^\dagger\delta(z).
\end{align}
Integrating across the junction, with inputs at $0+$ and outputs at $0-$, yields
\begin{align}
\hat c_{\rm out}&=\hat c_{\rm in}-\sqrt{2q}\,\hat u,\\
\hat d_{\rm out}&=\hat d_{\rm in}+\sqrt{2r}\,\hat v^\dagger.
\label{eq:complete_boundary}
\end{align}
The symmetric junction fields are consequently
\begin{align}
\hat c_0&=\hat c_{\rm in}-\sqrt{q/2}\,\hat u,\\
\hat d_0&=\hat d_{\rm in}+\sqrt{r/2}\,\hat v^\dagger.
\label{eq:complete_junction}
\end{align}
Since $[\hat u,\hat u^\dagger]=[\hat v,\hat v^\dagger]=0$ and $[\hat u,\hat v^\dagger]=[\hat v,\hat u^\dagger]=1$, the system Heisenberg equations before eliminating the local fields are
\begin{equation}
\dot{\hat u}=\sqrt{2r}\,\hat d_0^\dagger,
\qquad \dot{\hat v}=\sqrt{2q}\,\hat c_0.
\end{equation}
Substituting these junction fields and returning to $\hat a,\hat b^\dagger$ gives the Langevin equations
\begin{align}
\dot{\hat a}&=-\gamma\hat a+\gamma(1+\epsilon)\hat b^\dagger
 +\sqrt q\,\hat c_{\rm in}+\sqrt r\,\hat d_{\rm in}^\dagger,\\
\dot{\hat b}^\dagger&=\gamma\hat b^\dagger-\gamma(1+\epsilon)\hat a
 +\sqrt q\,\hat c_{\rm in}-\sqrt r\,\hat d_{\rm in}^\dagger.
\label{eq:complete_ab}
\end{align}
Compared with \eqref{eq:eom_minimal_cb1} and \eqref{eq:eom_minimal_cb2}, we have  an additional noise terms from $\hat d_{\rm in}$. 

Equivalently, we can rewrite the above equations of motion using $\hat u$ and $\hat v$ and the monitored output are
\begin{align}
\dot{\hat u}&=r\hat v+\sqrt{2r}\,\hat d_{\rm in}^\dagger,\\
\dot{\hat v}&=-q\hat u+\sqrt{2q}\,\hat c_{\rm in}\,.
\label{eq:complete_uvio}
\end{align}
At $\epsilon=0$ the second channel disappears and the protected common-reservoir model is recovered.

In the frequency domain, the solution of $\hat u$ away from the poles is
\begin{equation}
\hat u(\Omega)=\frac{r\sqrt{2q}\,\hat c_{\rm in}(\Omega)
-i\omega\sqrt{2r}\,\hat d_{\rm in}^\dagger(\Omega)}{D(\Omega)}.
\end{equation}
Here $\hat d_{\rm in}^\dagger(\Omega)$ denotes the Fourier transform of $\hat d_{\rm in}^\dagger(t)$, equivalently $[\hat d_{\rm in}(-\Omega)]^\dagger$, and we have defined 
\be
D(\Omega)\equiv \Omega_\epsilon^2-\Omega^2\,.
\ee 
The output of the readout mode follows form the input-output relation: $\hat c_{\rm out}=\hat c_{\rm in}-\sqrt{2q}\,\hat u$, which reduces to
\be
\hat c_{\rm out}=H(\Omega)\hat c_{\rm in}+K(\Omega)\hat d_{\rm in}^\dagger\,,
\ee
where the output transfer coefficients are
\begin{align}
H(\Omega)&=1-\frac{2\Omega_\epsilon^2}{D(\Omega)},\qquad
K(\Omega)=\frac{2i\Omega\Omega_\epsilon}{D(\Omega)}.
\label{eq:complete_transfer}
\end{align}
They obey $|H|^2-|K|^2=1$. With independent vacuum inputs, the two output quadratures have equal noise spectra and no anomalous field correlation; this realization is phase insensitive. In the single-sided vacuum convention, the phase quadrature of $\hat c_{\rm out}$ has the following noise  spectrum
\begin{align}
S^{\rm out}_{YY}(\Omega)
&=\left|1-\frac{2\Omega_\epsilon^2}{D(\Omega)}\right|^2
 +\frac{4\Omega^2\Omega_\epsilon^2}{D(\Omega)^2}\\
&=1+\frac{8\Omega^2\Omega_\epsilon^2}{(\Omega_\epsilon^2-\Omega^2)^2}.
\label{eq:complete_psd}
\end{align}
It recovers the scaling with respect to $\epsilon$ shown in Eq.\,\eqref{eq:static_output_noise} with an unimportant factor of 2 difference, when we consider the frequency sufficiently close to the EP. 

\bibliographystyle{apsrev4-2}
\bibliography{references}

\end{document}